\documentclass[nohyper]{JHEP}
\usepackage{cite}

\usepackage{epsfig}

\psfull
\newskip\humongous \humongous=0pt plus 1000pt minus 1000pt
\def\caja{\mathsurround=0pt}

\newif\ifdtup

\def\eqal2#1{\,\vcenter{\openup1\jot
\caja   \ialign{\strut \hfil$\displaystyle{##}$&\hfil$
\displaystyle{{}##}$\hfil &$
\displaystyle{{}##}$\hfil\crcr#1\crcr}}\,}

\def\fun#1#2{\lower3.6pt\vbox{\baselineskip0pt\lineskip.9pt
\newcommand{\half}{{\textstyle\frac{1}{2}}}
\newcommand{\ee}{e^+e^-}
\newcommand{\as}{\alpha_s}
\def\al{\alpha}
\newcommand{\be}{\beta}
\newcommand{\ce}{\langle C\rangle}
\def\eps{\epsilon}
\newcommand{\beq}{\begin{equation}}
\newcommand{\eeq}{\end{equation}}
\newcommand{\eq}[1]{(\ref{#1})}
\newcommand{\cl}[1]{{\cal #1}}
\newcommand{\rf}[1]{(\ref{#1})}
\newcommand{\sect}[1]{\section{#1}\setcounter{equation}{0}}
\newcommand{\secn}[1]{Section~\ref{#1}}
\newcommand{\nln}{\nonumber\\}
\newcommand{\mass}{{\mbox{\scriptsize mass}}}
\newcommand{\bub}{{\mbox{\scriptsize bub}}}
\renewcommand{\theequation}{\arabic{section}.\arabic{equation}}
\renewcommand{\labelenumi}{(\roman{enumi})}
\ialign{$\mathsurround=0pt#1\hfil##\hfil$\crcr#2\crcr\sim\crcr}}}
\def\cO#1{{\cal{O}}\left(#1\right)}

\def\re#1{(\ref{#1})}

\def\al{\alpha}
\def\be{\beta}

\def\cO{{\cal{O}}}

\def\half{\mbox{\small $\frac{1}{2}$}}

\def\as{\alpha_{{\textsc{s}}}}

\def\ee{e^+e^-}

\title{Universality of 1/Q corrections revisited}

\author{Mrinal Dasgupta \\
  Dipartimento di Fisica, Universit\`a di Milano Bicocca \\
  and INFN, Sezione di Milano,\\ 
  Via Celoria 16, I-20133, Milano, Italy \\
  E-mail: \email{dasgupta@mi.infn.it}}

\author{Lorenzo Magnea \\
 Dipartimento di Fisica Teorica, Universit\`a di Torino \\
 and INFN, Sezione di Torino, \\
 Via  P.~Giuria 1, I-10125, Torino, Italy \\
 E-mail: \email{magnea@to.infn.it}}

\author{Graham E. Smye 
    \thanks{Research supported by the U.K. Particle Physics
            and Astronomy Research Council.} \\
 Cavendish Laboratory, University of Cambridge, \\
 Madingley Road, Cambridge CB3 OHE, U.K. \\
 E-mail: \email{smye@hep.phy.cam.ac.uk}}  

\abstract{We provide an exact analytical calculation at the two--loop level
in the abelian limit of the leading power correction for the $C$ parameter 
in $e^{+}e^{-}$ annihilation. We compare our results to the numerical value 
obtained employing the soft approximation, the abelian part of the Milan 
factor. We demonstrate that a simple proportionality holds between the leading
power corrections to the $C$ parameter and to the longitudinal cross section 
in the soft region, and we verify that this proportionality holds for the 
full two--loop abelian contribution computed here. We comment on the 
possibility of extending this technique to other event shape variables and
distributions, as well as to the non--abelian contributions.}


\preprint{Bicocca--FT--99--34 \\ DFTT--56/99 \\
Cavendish--HEP--99/13 \\ hep-ph/9911316 \\ 
November 1999}

\begin{document}

\section{Introduction}
\label{intro}

The study of event shape variables in both $e^{+}e^{-}$ annihilation and 
DIS is at present one of the most active areas of phenomenological 
investigation in QCD. From a theoretical viewpoint existing predictions for 
the distributions of these variables emerge from the application of several
of the most advanced techniques of perturbative QCD, hence it is particularly 
satisfying to be able to confront theory successfully with experiment. 
Theoretical predictions for power corrections to shape variable 
distributions in fact require all order resummations of perturbation theory,
either within the framework of soft gluon resummation 
\cite{CTTW,CTW,CW,KS1,KS2}, or in the context of renormalon techniques 
\cite{BB,BBB,DW,AZ,BPY}, extended to account for the necessary two--loop 
effects \cite{BBM2,M1,M3,self}. 
In either case, the theoretical framework is adjusted for the inclusion 
of non--perturbative power corrections, using methods and models that may 
ultimately pave the way for a better understanding of low energy behaviour 
of QCD \cite{BPY,KS2}. On the experimental side, these variables have proved 
relatively simple to measure and have been one of the most popular sources 
for the extraction of the strong coupling from fits to the data 
\cite{ALEPH,DELPHI,OPAL,H1}.

In the present paper we will make use of the dispersive approach developed in
\cite{BB,BBB,DW,AZ,BPY}. In its original form, the dispersive approach was 
based on the resummation of fermionic bubble chains, a procedure which is 
strictly consistent only in the abelian limit, but can be reformulated under
rather general assumptions in terms of an effective low--energy QCD coupling,
for which a dispersive representation is adopted. Since the dispersive 
variable plays a role which is formally similar to a gluon mass, this 
approach is sometimes referred to as the ``massive gluon scheme''.
It was soon realized (first in a study by Nason and Seymour
\cite{NS}, who considered the thrust variable) that this method
could not directly be applied to general event shape variables, which are 
sensitive to the details of the decay process of the particular gluon being 
treated as ``massive''. A method to deal with the problem was proposed in
Refs.~\cite{BBM1,BBM2}, where it was shown that the bubble resummation could 
be performed without integrating inclusively over the gluon decay products,
and a calculation of the leading power corrections to the longitudinal cross 
section was performed in the abelian limit. A much more general analysis of 
the effects of non--inclusiveness on the most commonly used event shape 
variables was performed in Refs.\cite{M1,M3}, where the two--loop
correction due to gluon splitting was computed including non--abelian
terms. The main result of this detailed analysis is that the whole effect 
of non--inclusiveness, within the framework of the dispersive approach, is 
largely universal for the shape variables studied. Each observable acquires 
a two--loop enhancement factor to the naive ``massive gluon'' calculation,
and this enhancement factor, now called the Milan factor, is observable 
independent. All dependence on the chosen observable is encoded in the 
massive gluon result, so that the phenomenology is basically unaffected. 
Moreover, shape variables in DIS receive an identical enhancement \cite{self}.

The calculations leading to the Milan factor are done using soft two--loop 
matrix elements. This is justified because shape variables, being by 
construction essentially linear in particle momenta, are expected to 
receive leading ($1/Q$) power corrections only from soft gluon emission. 
In addition, soft kinematics is employed, where effects such as terms 
involving the square of the small transverse momentum of the gluon decay 
products are discarded, since they can only contribute at the 
level of $1/Q^2$ corrections. The universality of soft gluon radiation 
coupled with an underlying geometrical universality (linearity in emitted 
transverse momenta) in the shape variables themselves thus leads to the 
universal Milan factor appearing in every case.

In the present paper we test some of these assumptions by considering the 
$1/Q$ correction to two event shape variables, the $C$ parameter and the
longitudinal cross section. In the case of the $C$
parameter we perform the full calculation in the abelian (large $n_f$) limit,
in a manner analogous to the longitudinal cross section calculation of 
\cite{BBM2}, and we arrive at an analytical result for the 
$1/Q$ behaviour. The correction to the $C$ parameter was calculated in
the soft approximation in Ref.~\cite{M3}, and the Milan factor
enhancement obtained for it. The results we get here should be directly 
comparable to the abelian part of the Milan factor, since we allow only for 
gluon splitting into quarks.

We also show that a simple 
relationship holds between the $C$ parameter and the longitudinal cross 
section in the soft approximation. This relation is respected by the
$1/Q$ corrections computed here for $C$ and in reference \cite{BBM2}
for the longitudinal cross section, suggesting that the $1/Q$ 
corrections should be correctly obtained from the simplified soft approximation
advocated in the computation of the Milan factor.

We find however that the numerical 
result for the two--loop enhancement factor for the $C$-parameter in the 
abelian limit disagrees with the abelian limit of the Milan factor: if
we take the ratio of our result to the massive gluon one, we find an 
enhancement factor of $15 \pi^2/128 = 1.157$, whereas the corresponding
enhancement factor computed numerically in Ref.~\cite{M3} is of the
form $1+r_{ni}^{(a)} = 1.078$, where $r_{ni}^{(a)}$ is the ``non-inclusive'' 
piece of the Milan factor in the abelian limit. \footnote{To arrive at 
  this number for $r_{ni}^{(a)}$ we take $C_A$ = 0 in Eq.~(4.16) of
  Ref.~\cite{M1}, in the numerator as well as in $\beta_0$.}
In essence, the two results 
would agree if the correction factor $r_{ni}^{(a)}$ were doubled. 

The layout of the present paper is as follows. In Section $2$ we 
introduce the observables $C$ and $\sigma_L$, and we demonstrate 
their equivalence in the soft limit. In Section $3$ we show our 
calculation of the $C$ parameter in some detail. In the final section we 
compare our results to those obtained by other authors and briefly discuss
the different methods, as well as possible developments and applications.

\section{Definitions and the soft approximation}
\label{defin}

\subsection{$C$ parameter}

The $C$ parameter \cite{ERT} can be defined as 
\begin{equation}
C = 3 - \frac{3}{2} \sum_{i,j=1}^n 
\frac{(p_i\cdot p_j)^2}{(p_i\cdot q)(p_j\cdot q)}~~,
\label{cdef1}
\end{equation}
where $q^\mu$ is the photon four--momentum, $q^2 = Q^2$, and the sum is over 
all outgoing particles, so that each pair of particles is counted twice. 
An equivalent definition, valid in the $e^{+}e^{-}$ centre of mass frame, is
\begin{equation}
C = \frac{3}{2} \sum_{i,j=1}^n 
\frac{|\vec{p_i}||\vec{p_j}|}{Q^2} \sin^2 \theta_{ij}~~,
\label{cdef2}
\end{equation} 
where the sum once again runs over all outgoing particles. From these 
definitions, it is clear that the $C$ parameter vanishes at the 
Born level. The first contribution within perturbation theory appears at 
${\mathcal{O}}(\alpha_s)$  when one has a three body final state. 

\subsection{Longitudinal cross section}

The longitudinal cross section can be defined, within the context of single 
particle inclusive annihilation, by considering the differential cross section 
for the production of a hadron with given energy and angle with respect to the 
beam axis. The angular dependence of this cross section can be organized
in the form \cite{fragone}
\begin{equation}
\frac{d^2 \sigma^h}{d x d\cos \theta} = \frac{3}{8} 
\left(1 + \cos^2 \theta \right) \frac{d \sigma_T^h}{d x} + 
\frac{3}{4} \sin^2 \theta \frac{d \sigma_L^h}{d x} +
\frac{3}{4} \cos \theta \frac{d\sigma_A^h}{d x}~~,
\label{sigl}
\end{equation} 
with $x=2 p_h \cdot q/q^2$ being the hadron energy fraction and $\theta$ its 
angle with respect to the beam direction. The three terms on the right-hand 
side are referred to as the transverse, longitudinal and asymmetric 
contributions. The first two contributions respectively arise from 
gauge boson polarisation states transverse and longitudinal to the direction 
of the outgoing hadron $h$; the last term is a parity violating contribution 
that is absent in purely electromagnetic annihilation.

The longitudinal cross section itself is usually defined as the first moment
of $d\sigma_L^h/dx$, and it can be projected out from the single particle 
inclusive differential cross section in Eq.~\re{sigl} by multiplying with a 
suitable weight function and integrating over $\cos\theta$ and $x$, according 
to
\begin{equation}
\sigma_L = \frac{1}{2} \sum_h \int_0^1 x d x \int_{-1}^1 d\cos\theta \;
(2 - 5 \cos^2\theta)\frac{d^2 \sigma^h}{d x d \cos\theta}~~.
\label{sigl2}
\end{equation}
Note that since we are measuring an inclusive cross-section we can replace 
the hadronic sum above by the corresponding partonic one.

In the next subsection we shall make use of the above definitions to write 
down expressions for the longitudinal cross section and the $C$ parameter in 
the soft approximation, relevant to the computation of $1/Q$ corrections. 

\subsection{Soft approximation}

At the partonic level, in the soft approximation, the annihilation process 
produces a pair of hard fermions (``primary'' quark and antiquark), dressed
by soft gluon radiation. The primary quark and antiquark are essentially back 
to back and have an angular distribution about the beam axis that may be 
approximated by the purely transverse $1+\cos^2{\theta_0}$ pattern (the 
first term in Eq.~\re{sigl}), where $\theta_0$ is the quark angle with 
respect to the beam. In this approximation one is neglecting the recoil of 
the hard fermions due to soft gluon emission. 

Let us consider the contribution to $\sigma_L$ of a soft gluon, which is 
known to be the term responsible for the appearance of a $1/Q$ correction
\cite{BBM2,selffrag}. In order to do this, we need to introduce the gluon 
angle with respect to the beam $\theta_g$, while the gluon direction with 
respect to the parent quark may be specified by a polar angle $\theta$ and 
an azimuthal angle $\phi$. \footnote{An identical calculation relevant to 
the flux tube model of hadronisation was described by Nason and Webber
\cite{NW}.} In terms of these variables, the soft gluon contribution to the 
longitudinal cross section can be written as
\begin{eqnarray}
\label{sl}
\sigma_L^g & = & \frac{1}{2} \int_0^1 dx x \int_{-1}^1 d \cos \theta_g 
\; d \cos \theta_0 \; d \cos \theta \int_0^{2 \pi} \frac{d \phi}{2 \pi} 
\; \left(2 - 5 \cos^2 \theta_g \right) \nonumber \\
& \times & \frac{3}{8} (1 + \cos^2 \theta_0) 
\frac{d^2 \sigma^g}{d x d \cos\theta} \delta \left(\cos\theta_g - 
\cos\theta_0 \cos\theta + \sin\theta_0 \sin\theta \cos\phi \right) \\
& = & \frac{1}{4} \int_0^1 d x x \int_{-1}^{1} d\cos\theta \sin^2\theta 
\frac{d^2 \sigma^g}{d x d \cos\theta}~~, \nonumber
\end{eqnarray}  
where  $x$ is the gluon energy fraction, $x=2E_g/Q $, while here 
$d^2 \sigma^g/d x d \cos\theta$ is the differential cross section for soft 
gluon emission, given in the soft approximation by the standard 
$q\bar{q}$ antenna pattern.

The relationship with the $C$ parameter is easily established by noting that 
the only terms in the sum over outgoing particles that will contribute to 
the $1/Q$ correction for the mean value of $C$ are the terms pairing one of 
the primary fermions with the soft gluon. The contribution from the pairing
of the primary quark to the antiquark is $\cO(k_\perp^2)$, so that it can be 
discarded in the back--to--back (soft) limit. One finds then that in the 
soft approximation, concentrating on the terms responsible for $1/Q$ 
corrections, the mean value of $C$ can be calculated by weighting the matrix 
element $d^2\sigma^g/dx d\cos\theta$ with the function
\begin{equation}
C^{(1/Q)} = \frac{6}{Q^2} \frac{Q}{2} E_g \sin^2\theta = 
\frac{3}{2} x \sin^2\theta~~,
\label{softC}
\end{equation} 
which is precisely six times the weighting function for $\sigma_L$ as given in
Eq.~\re{sl}. One concludes that the power corrections to the expectation 
value of $C$ and to $\sigma_L$ are proportional, according to
\begin{equation}
\langle C \rangle^{(1/Q)} = 6 \; \sigma_L^{(1/Q)}~~.
\label{prop}
\end{equation}

In the next section we will calculate the exact value of the power correction 
for the $C$ parameter, taking proper account of gluon decay but in a 
large $n_f$ limit. We will then be able to compare with the calculation
of \cite{BBM2}, and verify that the $1/Q$ corrections to $\langle C 
\rangle$ and $\sigma_L$ indeed obey Eq.~\re{prop}.

\section{Calculations}

\label{calc}

In this section we provide some details of our calculation of the average
$\langle C \rangle$ due to gluon splitting into a (``secondary'') 
quark--antiquark pair. We will begin by performing the calculation in the 
naive massive gluon scheme, although such a calculation is manifestly 
incomplete, since we shall need it in order to compare our results with 
other available calculations.

\subsection{Massive gluon scheme}

In the dispersive approach (for a review, see \cite{benrep}), for a 
sufficiently inclusive observable, defined as an observable insensitive to
the momentum distribution of the gluon decay products, it is possible to 
classify power corrections by computing the observable at one loop with
Feynman rules appropriate to a massive gluon. One can then show
\cite{BB,BBB,DW,AZ,BPY} that power corrections are in one--to--one 
correspondence
with nonanalytic terms in the expansion of the result in powers of the 
squared gluon mass. It is easy to see that the $C$ parameter, like most 
event shape variables, is sensitive to the details of gluon decay, and thus
does not belong to the class of observables that can be dealt with using 
this method. One can however use the massive gluon result as a first estimate
of the size of the correction, and parametrize the full result as a 
(perturbative) enhancement of the massive gluon calculation. This is 
particularly useful since one can argue \cite{M1,M3} that all dependence 
on the particular observable at hand is contained in the massive gluon result.

Let us then consider a $q,\bar{q},g$ final state and let us label the 
four--momenta of the quark, antiquark and gluon by $p_1$, $p_2$ and $k$ 
respectively. The $C$ parameter assumes the simple form
\begin{equation}
C = 6 \frac{(1-x_1)(1-x_2)(1-x_3)}{x_1 x_2 x_3}~~,
\label{c3}
\end{equation}
where $x_{1,2} = 2 p_{1,2} \cdot q/q^2$ are the energy fractions carried by 
the quark and the antiquark respectively, while $x_3 = 2 k \cdot q/q^2$ is the 
gluon energy fraction. Energy conservation implies 
$\sum_i x_i = 2$. Notice that although we are doing a massive gluon 
calculation we have discarded the gluon mass in the definition of the event 
shape. This approximation has no effect on the coefficient of the $1/Q$ 
correction, in the present case; moreover, in order to compare our results with the Milan factor, 
we need to adopt the massless definition of $C$ in accordance with the 
procedure of Refs.~\cite{M1,M3}.

The expectation value of $C$ in the present approximation is given by
\begin{eqnarray}
\langle C \rangle_{(mg)} & = & \frac{1}{\sigma_0} \int d \sigma 
\; C \nonumber \\ & = & \frac{\alpha_s}{16 \pi N_c} \int_0^{1 - \epsilon} 
d x_1 \int_{1 - x_1 - \epsilon}^{1 - \frac{\epsilon}{1 - x_1}} d x_2 \; 
W^{\mu\al} W_{\mu\al}^* \; C~~,
\label{Cint}
\end{eqnarray}
where $\sigma_0$ is the Born cross section, $\epsilon$ is the square of the 
gluon mass divided by $q^2$, and $(-ie)(-ig)W^{\mu\al}$ is the matrix element 
for the decay of the virtual photon (with polarisation index $\mu$) into a 
quark, an antiquark, and a gluon with polarisation index $\al$. Current 
conservation then implies that $q_\mu W^{\mu\al} = 0$ and $k_{\al} W^{\mu\al} 
= 0$. In terms of the energy fractions $x_i$,
\begin{equation}
W^{\mu\al} W_{\mu\al}^* = 8 N_c C_F \left[ 
\frac{x_1^2 + x_2^2 + 2 \epsilon (x_1 + x_2 + \epsilon)}{(1 - x_1)(1 - x_2)}
- \frac{\epsilon}{(1 - x_1)^2} - \frac{\epsilon}{(1 - x_2)^2} \right]~~.
\label{WW}
\end{equation}

The integrals in Eq.~\re{Cint} are easily computed. Expanding the result in 
powers of $\epsilon$ we get
\begin{equation}
\langle C \rangle_{(mg)} = \frac{C_F}{2\pi} \alpha_s 
\left[4 \pi^2 - 33 - 12 \pi \sqrt{\epsilon} + \cO(\epsilon) \right]~~.
\label{cmg}
\end{equation}
In the language of the dispersive approach, this translates into a power 
correction, generated by the $\sqrt{\epsilon}$ term, given by (see \cite{BPY})
\begin{equation}
\langle C \rangle_{(mg)}^{1/Q} = 6 \pi \frac{A_1}{Q}~~,
\label{cmgpow}
\end{equation}
where the dimensional parameter $A_1$ is interpreted as the first (half 
integer) moment of the non--perturbative component of the effective coupling, 
as defined in \cite{BPY},
\begin{equation}
A_1 = \frac{C_F}{2 \pi} \int_0^\infty \frac{d \mu^2}{\mu^2} \;\mu \; 
\delta \alpha_{{\mbox{\tiny{eff}}}}(\mu^2)~~.
\label{halfmom}
\end{equation}
A completely equivalent expression for the power correction follows from a 
purely perturbative analysis, as a consequence of the sum over ``bubble'' 
graphs (see \cite{BB,BBB}). In that case the coefficient $A_1$ is related to
a slightly different effective coupling function, related to the present 
one essentially by a change of renormalization scheme (for a comparison of
the two points of view, see \cite{benrep}).

\subsection{$C$ parameter with gluon splitting}

Let us now consider the splitting of the gluon into a quark--antiquark pair in 
more detail. We focus on a four--particle final state with a ``primary''
$q \bar{q}$ pair, carrying momenta $p_1$ and $p_2$, while the secondary quark 
and antiquark have momenta $k_1$ and $k_2$. Summing the relevant graphs, and 
summing in each graph over all insertions of fermion bubbles in the gluon 
propagator, we are led to replace Eq.~\re{Cint} by
\begin{eqnarray}
\langle C \rangle & = &  
\frac{3}{N_c} \left(\frac{2 \pi}{Q^2}\right)^2 \alpha_s^2 T_R n_f 
\int \frac{d k^2}{k^4} \frac{1}{\left| 1 + \Pi(k^2) \right|^2}
\int\! d\,{\rm Lips}[q \to p_1,p_2,k] \,
W_{\mu \al} \, W_{\nu \beta}^* \, L^{\mu \nu} \nonumber \\
& \times & \int d\,{\rm Lips}[k \to k_1,k_2]
\; {\rm Tr}[\gamma^\al\!\not\!k_1\gamma^{\beta}\!\not\!k_2]
\, C~~.
\label{cab}
\end{eqnarray}
Here $d\,{\rm Lips}[q \to \{p_i\}]$ is the appropriate Lorentz--invariant 
phase space measure, $\Pi(k^2)$ is the renormalized one--loop gluon vacuum 
polarization induced by quarks, and $L^{\mu \nu}$ is the leptonic tensor, 
which will be substituted by the corresponding average over the beam 
orientation, $\langle L^{\mu \nu} \rangle = - 4 Q^2 g^{\mu \nu}/3$, 
as customary when working with event shape variables. 
In Eq.~\re{cab} we have factorized the four--particle 
phase space to introduce an explicit integral over the square 
of the gluon four--momentum, $k^2$, which of course plays the role of the 
dispersive variable (``gluon mass'') in the present calculation. Thus we will
be interested in the $k^2$--dependent integrand of Eq.~\re{cab}, and in 
particular in its expansion in powers of $\sqrt{k^2}$. Notice that a factor
of $1/(1 + \Pi(k^2))$ is just what is needed to turn the (fixed) coupling 
$\alpha_s$ into the running coupling evaluated at scale $k^2$, in the abelian 
limit. For $C$ we take the full expression for 4 outgoing particles, which 
we write as
\begin{equation}
C = C_{(p)} + C_{(m)} + C_{(s)}~~,
\label{splitc}
\end{equation}
where $C_{(p)}$ is the ``inclusive'' term, involving only the momenta of the 
primary fermions,
\begin{equation}
C_{(p)} = 3 - 3 \frac{(p_1 \cdot p_2)^2}{(p_1 \cdot q)(p_2 \cdot q)}~~,
\label{cp}
\end{equation}
$C_{(s)}$ is the term involving only the momenta of the secondary fermions,
\begin{equation}
C_{(s)} = - 3 \frac{(k_1 \cdot k_2)^2}{(k_1 \cdot q)(k_2 \cdot q)}~~,
\label{cs}
\end{equation}
and finally $C_{(m)}$ is the sum of all mixed terms. Using the symmetries of
the integral in Eq.~\re{cab} we can freely replace $k_2$ with $k_1$ and $p_2$
with $p_1$ in $C_{(m)}$, thus we can use
\begin{equation}
C_{(m)} = - 12 \frac{(p_1 \cdot k_1)^2}{(p_1 \cdot q)(k_1 \cdot q)}~~.
\label{cm}
\end{equation}
Since we are interested in the distribution of $C$ as a function of the 
gluon ``mass'' $k^2$, for each of the three contributions to $\langle C 
\rangle $ we write
\begin{equation}
\langle C \rangle_{(i)} = 
\int \frac{d \epsilon}{\epsilon} \frac{1}{\left| 1 + \Pi(Q^2 \epsilon) 
\right|^2} \; {\cal C}_{(i)}(\epsilon)~~,
\label{cdens}
\end{equation}
where $i = p,m,s$ and $\epsilon = k^2/Q^2$. Our task is to compute the
distributions ${\cal C}_{(i)}(\epsilon)$, and extract the non--analytic 
behaviour for small values of $\epsilon$.

The primary contribution to $C$ is clearly the easiest to evaluate, since the
integration over $k_1$ and $k_2$ can be done inclusively. Using the 
transversality of the hadronic tensor, $k_{\al} W^{\mu\al} = 0$, one can 
simply substitute for the bubble integral
\begin{equation}
\int d\,{\rm Lips}[k \to k_1,k_2]
\; {\rm Tr}[\gamma_\al\!\not\!k_1\gamma_{\beta}\!\not\!k_2] \to
- \frac{1}{6 \pi} \, Q^2 \, \epsilon \, g_{\al \beta}~~.
\label{ibubb}
\end{equation}
One finds then
\begin{equation}
{\cal C}_{(p)}(\epsilon) = \frac{1}{N_c} \frac{\alpha_s^2 T_R n_f}{48 \pi^2}
\int_0^{1 - \epsilon} d x_1
\int_{1 - x_1 - \epsilon}^{1 - \frac{\epsilon}{1 - x_1}} d x_2 \; 
W^{\mu\al} W_{\mu\al}^* \; C_{(p)}~~.
\label{cpdens}
\end{equation}
The remaining integrals are easily performed, and expanding around
$\epsilon = 0$ yields
\begin{equation}
{\cal C}_{(p)}(\epsilon) = \frac{C_F}{2 \pi} \frac{\alpha_s^2 T_R n_f}{3 \pi}
\left[- 8 \ln \epsilon - \frac{133}{6} + \cO(\epsilon) \right]~~.
\label{cpfin}
\end{equation}
Notice that ${\cal C}_{(p)}(\epsilon)$ by itself is infrared divergent. The
divergence will be cancelled by the mixed contribution, to which we now turn.

The $\epsilon$--distribution arising from the mixed terms is by far the most 
difficult to evaluate analytically. It is given by
\begin{equation}
{\cal C}_{(m)}(\epsilon) =
\frac{3}{N_c} \frac{\alpha_s^2 T_R n_f}{2 \pi} \frac{1}{Q^2 \epsilon} 
\int_0^{1 - \epsilon} dx_1 
\int_{1 - x_1 - \epsilon}^{1 - \frac{\epsilon}{1 - x_1}} dx_2 \; 
\frac{1}{p_1 \cdot q} W^{\mu\alpha} W_\mu^{*\beta} T_{\alpha\beta}~~,
\label{cmdens}
\end{equation}
where
\begin{equation}
\label{Tdef}
T_{\alpha\beta} = \int d\,{\rm Lips}[k \to k_1,k_2]
\; {\rm Tr}[\gamma_\al\!\not\!k_1\gamma_{\beta}\!\not\!k_2]
\frac{(p_1\cdot k_1)^2}{k_1\cdot q}~~.
\end{equation}
There are several ways to evaluate the bubble integral $T_{\alpha\beta}$. 
Perhaps the most straightforward is to note that $T_{\alpha \beta}$ is 
symmetric, and obeys $k^{\alpha}T_{\alpha \beta}=0$. Then we can write the 
decomposition
\begin{eqnarray}
T_{\alpha\beta} & = & A_1 \, Q^2 \left(g_{\alpha\beta} - 
\frac{k_\alpha k_\beta}{k^2} \right) + A_2 \left(p_{1\alpha} - 
\frac{p_1 \cdot k}{k^2} k_\alpha \right) \left(p_{1\beta} - 
\frac{p_1 \cdot k}{k^2} k_\beta  \right) \nonumber \\
& + & A_3 \left[ \left( p_{1\alpha} - \frac{p_1\cdot k}{k^2} k_\alpha \right)
\left( q_\beta - \frac{q\cdot k}{k^2} k_\beta \right) + 
\left(\alpha \leftrightarrow \beta \right) \right] \nonumber \\
& + & A_4 \left( q_\alpha - \frac{q\cdot k}{k^2} k_\alpha \right)
\left( q_\beta - \frac{q\cdot k}{k^2} k_\beta \right)~~,
\label{decomp}
\end{eqnarray}
where the $A_i$ are scalar functions of $p_1,k$ and $q$, which can be 
evaluated, for example, by integrating Eq.~\re{Tdef} component by 
component in the rest frame of $k$.

It follows from Eqs.~\re{cmdens} and \re{decomp}, and from current 
conservation, that only four independent projections of the tensor 
$W^{\mu\alpha}W_{\mu}^{*\beta}$ are needed. The projection with the metric 
tensor $g_{\alpha\beta}$ gives Eq.~\re{WW}; the other projections are 
given by
\begin{eqnarray}
p_{1\alpha} p_{1\beta} W^{\mu\alpha} W_\mu^{*\beta} & = & 4 N_c C_F Q^2 
\frac{x_1^2 (x_3 - 1 - \epsilon)}{(1 - x_1)^2} \nonumber \\
(p_{1\alpha} q_\beta + q_\alpha p_{1\beta}) W^{\mu\alpha} W_\mu^{*\beta} & = & 
8 N_c C_F Q^2 \left[\frac{x_1^2 (x_3 - 1 - \epsilon)}{(1 - x_1)^2} + 
\frac{(x_3 - 1 - \epsilon)^2}{(1 - x_1)(1 - x_2)} \right] \nonumber \\
q_\alpha q_\beta W^{\mu\alpha} W_\mu^{*\beta} & = & 4 N_c C_F Q^2 \left[ 
\frac{x_1^2 (x_3 - 1 - \epsilon)}{(1 - x_1)^2} + 
\frac{2 (x_3 - 1 - \epsilon)^2}{(1 - x_1)(1 - x_2)} \right. \nonumber \\
& & + \left. \frac{x_2^2 (x_3 - 1 - \epsilon)}{(1 - x_2)^2} \right]~~.
\label{proj}
\end{eqnarray}
To simplify the computation, a convenient change of variables is
\begin{eqnarray}
u & = & x_1 - x_2 \\
v + \frac{\epsilon}{v} & = & 2 - x_1 - x_2 = x_3~~.
\end{eqnarray}
In terms of these variables, the scalars entering the decomposition of the
tensor $T_{\alpha\beta}$, Eq.~\re{decomp}, have the general form
\begin{equation}
\frac{Q^2}{p_1 \cdot q} A_i = \frac{P_i(u,v,\epsilon) + 
Q_i(u,v,\epsilon) \ln v + R_i(u,v,\epsilon) \ln \epsilon}{(v^2 - 
\epsilon)^9(v^2 - 2v - uv + \epsilon)}
\label{scala}
\end{equation}
where $P_i$, $Q_i$ and $R_i$ are polynomial functions of $u$, $v$ and 
$\epsilon$, whereas the projections of the hadronic tensor given in 
Eq.~\re{proj} are rational functions of the same variables. All the integrals
involved in the computation of Eq.~\re{cmdens} can be performed analytically,
through a rather lengthy process of partial fractioning. The result is 
expressible in terms of di-- and trilogarithms of functions of $\epsilon$,
and can be expanded around $\epsilon = 0$ yielding
\begin{equation}
{\cal C}_{(m)}(\epsilon) = \frac{C_F}{2\pi} \frac{\alpha_s^2 T_R n_f}{3\pi}
\left[ 8 \ln \epsilon + 4 \pi^2 - \frac{65}{6} - 
\frac{45 \pi^3}{32} \sqrt{\epsilon} + \cO(\epsilon) \right]~~.
\label{cmfin}
\end{equation}
As promised, the infrared divergence (as $\epsilon \to 0$) of Eq.~\re{cmfin}
cancels the one of Eq.~\re{cpfin}.

The final contribution to $\langle C \rangle$ is that from the secondary 
terms, arising from the distribution ${\cal C}_{(s)}(\epsilon)$. 
By inspection, this contribution must be at least $\cO(\epsilon)$, since 
the corresponding event shape, Eq.~\re{cs} is quadratic in the gluon energy
in the soft region, and thus cannot contribute to the leading power 
correction. This can be confirmed by an explicit calculation along the lines
of the ones leading to Eqs.~\re{cpfin} and \re{cmfin}.

Our final result for the $C$ parameter $\epsilon$ distribution is
\begin{equation}
{\cal C}(\epsilon) = \frac{C_F}{2\pi} \frac{\alpha_s^2 T_R n_f}{3\pi}
\left[ 4 \pi^2 - 33 - \frac{45 \pi^3}{32} \sqrt{\epsilon} + \cO(\epsilon) 
\right]~~.
\label{cfin}
\end{equation}
Following \cite{BB,BBB,BBM2}, one can directly compare Eq.~\re{cfin} with
Eq.~\re{cmg}, upon noticing that the coefficient $\alpha_s^2 T_R 
n_f/(3 \pi)$ should be read as $\alpha_s^2 \beta_0^f$, where $\beta_0^f$ is the
fermion contribution to the one--loop $\beta$ function. 
From here the power correction can be obtained by using the
renormalisation group equation for the running coupling to replace the 
``spectral function'', $\beta_0^f \alpha_s^2(k^2)$, by the logarithmic
derivative of the coupling with respect to its argument and
integrating by parts.  The coefficient
of $\sqrt{\epsilon}$ is mapped to the coefficient of the $1/Q$ correction
as in Eqs.~\re{cmg} and \re{cmgpow}.

Using the language of Refs.~\cite{M1,M3}, one would interpret Eq.~\re{cfin} as the sum of the
``naive'' contribution to the series of power corrections plus the fermionic 
part of the ``non inclusive'' correction to it. We find that the exact calculation
leading to Eq.~\re{cfin} gives a $1/Q$ correction larger than the one computed
in the massive gluon scheme by a factor $15 \pi^2/128 = 1.157$, as announced 
in the introduction.

\subsection{Comparison with $\sigma_L$}

A calculation analogous to the one just outlined was performed for the 
longitudinal cross section in Ref.~\cite{BBM2}, where a model for the 
distribution in energy fraction of the power correction was also constructed.
It was also noted there that the $1/Q$ correction to $\sigma_L$ arises from
a resummation of power corrections of the form $1/(x Q)^{2 n}$ in the 
distributions, which have no correction with odd powers of
$1/Q$; this was also the conclusion reached in \cite{selffrag}, within 
the massive gluon model.
This shows
that the leading power correction to the total $\sigma_L$ is entirely due 
to the soft gluon region, $x \to 0$. The result given in Ref.~\cite{BBM2}
for $\sigma_L$ can be written in the form
\begin{equation}
\sigma_L(\epsilon) = \frac{C_F}{2\pi} \frac{\alpha_s^2 T_R n_f}{3\pi}
\left[ 1 - \frac{15 \pi^3}{64} \sqrt{\epsilon} + \cO(\epsilon) 
\right]~~.
\label{sigfin}
\end{equation}
One sees that the two results indeed satisfy Eq.~\re{prop}, which was 
obtained in the soft approximation. This confirms that the leading power 
correction arises entirely from the emission of soft gluons, and from their 
subsequent splitting.

\section{Discussion}

We performed an analytical calculation, with two--loop accuracy and in the 
abelian limit, of the leading power correction to the $C$ parameter measured 
in $e^+ e^-$ annihilation; we showed that our result is simply related to 
the corresponding result for the longitudinal cross section, and we 
compared it to existing calculations. We explicitly checked that the entire power 
correction comes from the region in which the splitting gluon is soft: this 
can be shown by tracking the contributions to the final answer through the 
calculation, and is confirmed by the fact that the simple relationship we 
find between $\langle C \rangle$ and $\sigma_L$ is a property of the soft 
approximation. Our calculation is thus a positive test of the applicability 
of the soft approximation to the computation of $1/Q$ corrections to event
shape observables. We find however a numerical discrepancy with the results
of \cite{M1,M3}, which we should reproduce in the abelian limit.

Our technique, in essence a straightforward if lengthy two--loop calculation,
is applicable in principle to all event shape variables. Although in some 
cases it may turn out to be too cumbersome to generate the full analytical 
result, it is always possible, and in fact simple, to produce a 
one--dimensional integral representation for the answer, from which the 
coefficient of the desired power correction can always be derived numerically.
The fact that the calculation is performed in the abelian limit is not a 
severe problem, it should rather be viewed as a technical simplification. 

Corrections to the abelian limit come from two sources: on the one hand,
a subset of the non--abelian diagrams serves to reconstruct the full one--loop
$\beta$ function, $\beta_0$, from its abelian counterpart, $\beta_0^f$; on the 
other hand, the fact that the event shape is sensitive to the details of 
gluon splitting generates contribution that are not proportional to $\beta_0$,
since they are not directly related to the running of the coupling.
The first source of corrections (which must be taken into account also in 
the simplifed massive gluon scheme) is under control: there are strong 
physical arguments for such corrections to be there, and furthermore, at 
least in principle, they could be explicitly included by using existing
techniques to isolate gauge--invariant subsets of diagrams such as the 
pinch technique \cite{papa}. The second set of corrections is related to 
the splitting of the gluon into two gluons, and these are the corrections 
included in the soft approximation in the Milan factor. If it were to prove
impossible to iron out the numerical differences between the two approaches,
an analytic computation of non--abelian gluon splitting can in principle be 
performed. 

A more interesting issue from a phenomenological point of view is
the study of event shape distributions, such as $d \sigma/d C$ or $d \sigma_L
/ d x$, with $x$ the energy fraction of the detected hadron. Such studies can 
be performed with the technique outlined here, and in fact in \cite{BBM2} the
$\sigma_L$ distribution was studied, showing how the $1/Q$ correction to 
$\sigma_L$ arises from a summation of even power correction to all orders.
However, having shown that the power correction to the average event shape 
is determined by soft gluon emission, we expect to recover at least 
qualitatively the results of \cite{M1,M3}, namely the characteristic
constant shift in the distribution from its perturbative estimate by a power 
suppressed amount. It would be interesting to understand to what extent this 
simple behaviour of the distributions is due to the approximations inherent in 
the dispersive approach. In fact the shift in the distributions is recovered
in a factorization--based approach \cite{KS2} only as an approximation of 
the full answer, obtained essentially by neglecting the long--range, 
wide--angle correlations of the soft radiation. A complete description of the 
power correction to the distribution requires in that approach a 
non--perturbative function rather than a single parameter. 
The phenomenological impact of a more detailed analysis of the power 
correction along these lines is at present an open question.

\medskip
\medskip
\medskip

\noindent{\large\bf Acknowledgements}
We would like to thank Yuri Dokshitzer, Giuseppe Marchesini, Gavin Salam, 
Bryan Webber and Joannis Papavassiliou for helpful discussions and 
clarifications. One of us (MD) would like to acknowledge the hospitality of 
the HEP group at the Cavendish Laboratory in Cambridge and of the INFN 
Sezione di Torino where part of this work was carried out. 
The work was supported in part by the EU Fourth Framework Programme 
`Training and Mobility of Researchers', Network `Quantum Chromodynamics and 
the Deep Structure of Elementary Particles', contract FMRX-CT98-0194 
(DG 12 - MIHT).

\medskip
\medskip
\medskip
\noindent{\large\bf Note}
After the discrepancy between our results and those of
Refs.~\cite{M1,M3} had become evident, an error of a factor of two was found
in the computation of the Milan factor
\cite{YuBlois}. Once this error is corrected, there is complete
agreement with our calculation. Moreover it becomes apparent that 
$\sigma_L$ indeed belongs to the family of
observables that receive the universal Milan enhancement, there no longer
being any conflict with Ref.~\cite{BBM2}. This should enable
experimental investigation of the power correction to $\sigma_L$ just as for
other shape variables.

\end{document}